\documentclass[aps,prl,twocolumn,superscriptaddress,showpacs,floatfix]{revtex4}
\usepackage{graphicx}

\newcommand{\eq}[1]{Eq.~(\ref{#1})}
\newcommand{\fig}[1]{Fig.~\ref{#1}}
\newcommand{\be}[1]{\begin{equation}\label{#1}}
\newcommand{\ee}{\end{equation}}

\begin{document}

\title{Antiblockade in Rydberg excitation of an ultracold lattice gas}

\author{C.\ Ates}
\affiliation{Max Planck Institute for the Physics of Complex Systems, N{\"o}thnitzer
Str.\ 38, D-01187 Dresden, Germany}
\author{T.\ Pohl}
\affiliation{ITAMP, Harvard-Smithsonian Center for Astrophysics, 60 Garden
Street, Cambridge, MA 02138, USA}
\author{T.\ Pattard}
\affiliation{Max Planck Institute for the Physics of Complex Systems, N{\"o}thnitzer
Str.\ 38, D-01187 Dresden, Germany}
\author{J.\ M.\ Rost}
\affiliation{Max Planck Institute for the Physics of Complex Systems, N{\"o}thnitzer
Str.\ 38, D-01187 Dresden, Germany}

\date{\today}

\begin{abstract}
    It is shown that the two-step excitation scheme typically used to
    create an ultracold Rydberg gas can be described with an effective
    two-level rate equation, greatly reducing the complexity of the
    optical Bloch equations.  This allows us to solve the many-body
    problem of interacting cold atoms with a Monte Carlo technique.  Our
    results reproduce the Rydberg blockade effect.  However, we 
    demonstrate that
    an Autler-Townes double peak structure in the two-step excitation
    scheme, which occurs for moderate pulse lengths as used in the
    experiment, can give rise to an antiblockade effect. It is observable
    in a lattice gas with regularly spaced atoms.  Since the
    antiblockade effect is robust against a large number of lattice
    defects it should be experimentally realizable with an optical
    lattice created by CO$_{2}$ lasers.
\end{abstract}

\maketitle

The prediction of a blockade effect in
the Rydberg excitation of ultracold gases  due to long-range 
interaction \cite{lufl+01} has sparked
several experiments which have demonstrated the blockade effect under
different circumstances.  In \cite{tofa+04} Rydberg excitation to n =
30--80 was achieved by a single photon of a pulsed UV laser from the
5s Rb ground state.  In \cite{sire+04,lire+05} a two-step
scheme is used where a strong pump laser drives the 5s-5p transition
of Rb atoms while a tunable second laser excites from 5p to a
Rydberg $n\ell$ state.  In all three cases a suppression  of the
excitation has been observed as a function of increasing laser
intensity or/and density of the gas, i.e., as a function of increasing
effective interaction between the atoms which shifts the energy levels
out of resonance with the laser light.

\begin{figure}
    \includegraphics[width=0.55\columnwidth]{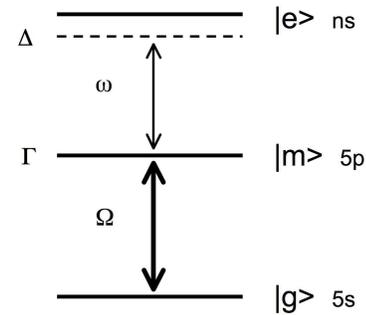}
    \caption{Sketch of the two-step excitation scheme.}
    \label{fig:sketch}
\end{figure}

However, the two different excitation schemes (single UV pulse and two-step
excitation) have a very different nature and may lead to
dramatic differences in the blockade behavior.  In fact, as we will
show below, the two-step scheme (see \fig{fig:sketch}) may even induce
an antiblockade effect due to the form of the single Rydberg atom 
population  $P_{e}(t,\Delta)$ as a function of the detuning
$\Delta$ from resonant excitation.  This antiblockade effect will
only be uncovered by a structured ultracold gas, e.g., a lattice gas
with atoms regularly placed on a 3-dimensional lattice.  Yet, the
condition on the regular arrangement is not too strict: even with 20
\% lattice defects the antiblockade is still visible.

The key to the antiblockade effect is the excitation dynamics of the
final Rydberg state $n\ell$ in a three-level system with the
transition between level $|g\rangle$ (ground state) and level
$|m\rangle$ (intermediate state) driven by a strong optical transition
with Rabi frequency $\Omega \gg \omega$, the Rabi frequency which
drives the transition between level $|m\rangle$ and the Rydberg level
$|e\rangle$, see \fig{fig:sketch}.

Experimentally, the intermediate level $|m\rangle$
decays with a rate $\Gamma\gg\omega$,  large compared to 
the upper Rabi frequency $\omega$.  
Under the conditions $\Omega \gg \omega$ and $\Gamma\gg\omega$
and a sufficiently long
laser pulse the optical Bloch equations \cite{wist76} for transitions in this 
three-level system reduce to a rate equation for a two-level system which
contains the upper state $|e\rangle$ and an effective lower state 
$|\bar g\rangle$
\cite{ates*long+06}. The reason for this
simplification lies in the strong decay $\Gamma$ which damps out
coherences relatively fast and  ultimately allows for the rate
description which directly gives the probability $P_{e}$ for an atom
to be in the upper Rydberg state $|e\rangle$, 
\be{reexc}
P_{e}(t,\Delta) = 
P_{\infty}(\Delta)\left(1-\exp
\left[-\frac{\gamma(\Delta)t}{P_{\infty}(\Delta)}\right]\right)\,, 
\ee
where $P_{\infty}=P_{e}(t\to\infty,\Delta)$ is the steady-state occupation of
$|e\rangle$ and $\gamma(\Delta)$ is the rate of populating $|e\rangle$ for short 
times $t$.

\begin{figure*}
     \includegraphics[width=1.0\columnwidth]{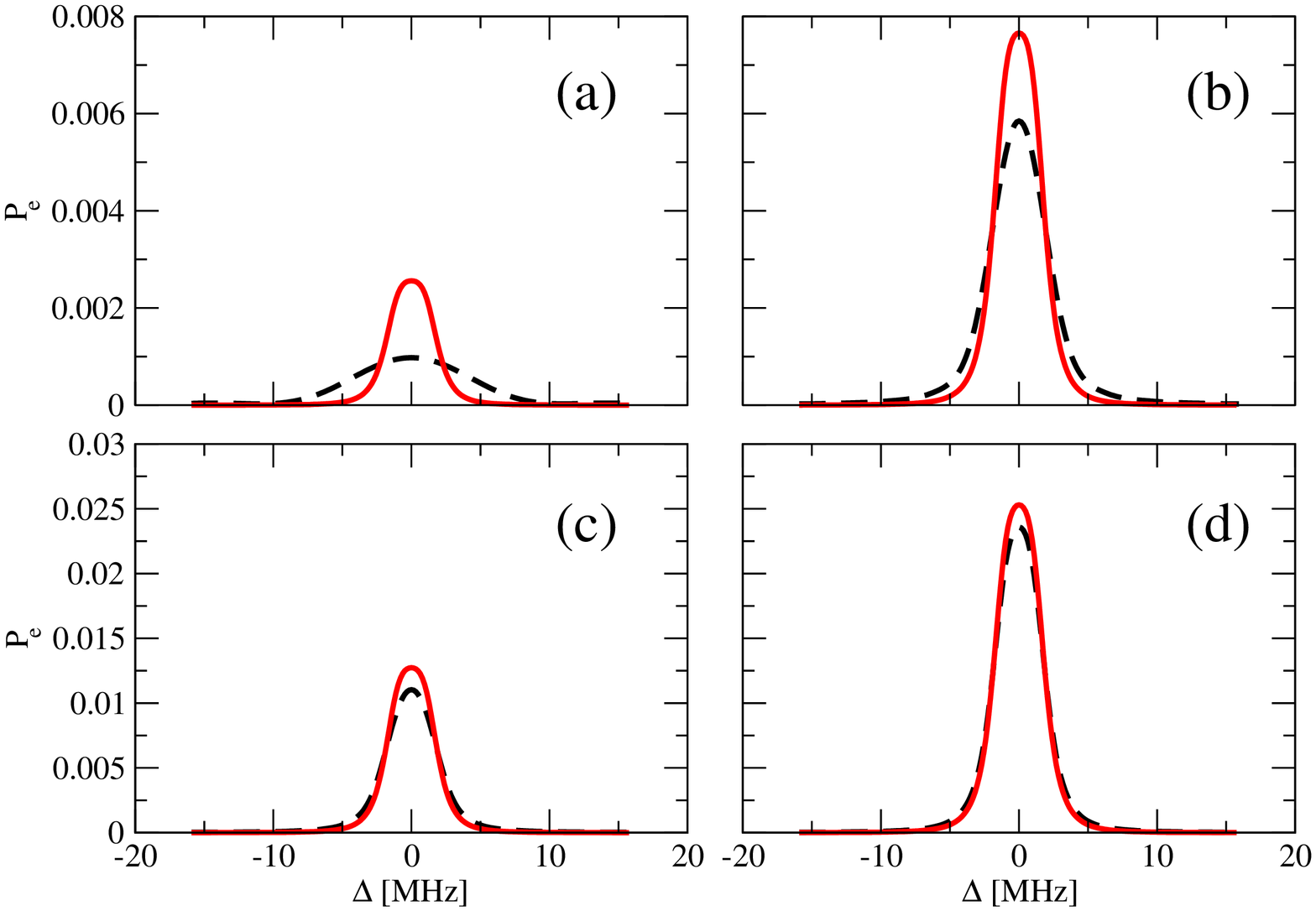}\hfill
     \includegraphics[width=1.0\columnwidth]{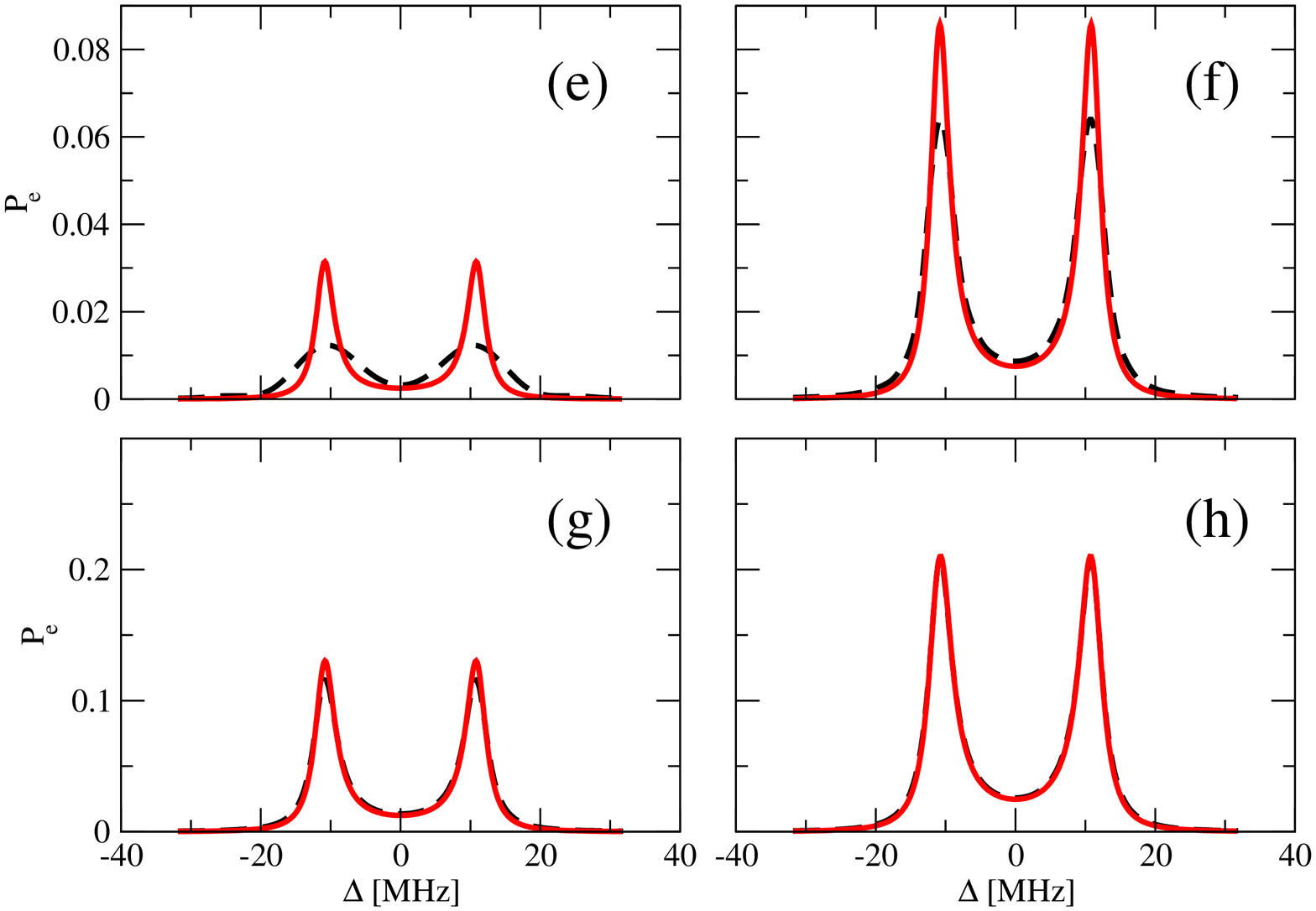}
     \caption{The population $P_{e}$ of the Rydberg state $|e\rangle$ in the 
     three-level system of
     \fig{fig:sketch} according to the rate equation \eq{reexc} 
     (solid) 
     and the optical Bloch equation (dashed)  for laser pulse lengths 
     of 0.1$\mu$s (a,e), 0.3$\mu$s (b,f),  0.5$\mu$s (c,g), 
     1.0$\mu$s (d,h).
     The parameters are  
      $\Omega=2\pi\cdot 4\,$MHz, $\omega=2\pi\cdot 0.2\,$MHz, 
      $\Gamma = 2\pi\cdot 6\,$MHz  for the left set (a-d) and $\Omega=2\pi\cdot 
      22.1\,$MHz, $\omega=2\pi\cdot 
 	0.8\,$MHz, $\Gamma = 2\pi\cdot 6\,$MHz for the right set 
	(e-h).
      }
      \label{fig:noDP}
 \end{figure*}

Typical Rydberg populations as a function of detuning $\Delta$ are
shown in the left part of \fig{fig:noDP} which also demonstrates that, for the given
Rabi frequencies and decay rate, a pulse length of $t \ge 0.5\,\mu$s
is enough to make the description with a rate equation applicable.

For atoms embedded in an environment of other atoms, e.g., a gas, the 
Rydberg level $|e\rangle$ of a ground state atom is shifted by 
$\delta$ due to the (weak) interaction 
with Rydberg atoms in the neighborhood.

  We use the simple picture as formulated in \cite{lita+05} for Rb.
 A pair of Rydberg atoms in state $ab$ at distance $R$ experiences a shift 
 $\delta(R)$ of its electronic energy due to an induced dipole 
 coupling  $V(R) = \mu_{aa'}\mu_{bb'}/R^{3}$ to an energetically close 
 state $a'b'$. It is given by the eigenvalues
 \be{shift}
 \delta(R) = \textstyle\frac 
 12(\delta_{0}\pm(\delta_{0}^{2}+4V^{2})^{\frac 12})
 \ee
 of the two-state matrix Hamiltonian 
 with elements
 $H_{11}=0$, $H_{22}=\delta_{0}$, $H_{12}=H_{21}=V(R)$,
 where $\delta_{0}$ is the asymptotic ($R\to\infty$) difference 
 between the energies of the two pairs.
 
The relevant neighboring pair to an excited pair $ab = nsns$,
corresponding to two atoms in state $|e\rangle$ of \fig{fig:sketch}, is
$a'b'=(n-1)p_{3/2}np_{3/2}$. For a specific quantum number $n_{0}$ we may define
$\mu^{2}(n_{0})\equiv \mu_{n_{0}s(n_{0}-1)p}\mu_{n_{0}sn_{0}p}$.  We
have taken its value and the value for $\delta_{0}$ 
($\mu^{2}(n_{0})=843800\,$a.u., $\delta_{0}=
-0.0378\,$a.u.\ for $n_{0}=48$) from \cite{lita+05} and 
adapted
to our Rydberg levels by appropriate scaling in $n$ \cite{ga94},
\begin{eqnarray}\label{n-scale}
\mu^{2}(n)&=&\mu^{2}(n_{0})
\left(\frac{n^{*}}{n^{*}_{0}}\right)^{4}\nonumber\\
\delta_{0}(n) &=& 
\delta_{0}(n_{0})\left(\frac{n^{*}_{0}}{n^{*}}\right)^{3}\,,
\end{eqnarray}
where $n^{*}=n-\eta$ includes the appropriate quantum defect, for 
the $ns$ states of Rb $\eta = 3.13$. 
Furthermore, we
excite the repulsive branch of \eq{shift} which finally   defines the shift
$\delta$ resulting from  a single  excited atom.  

\begin{figure}[hb]
       \includegraphics[width=0.8\columnwidth]{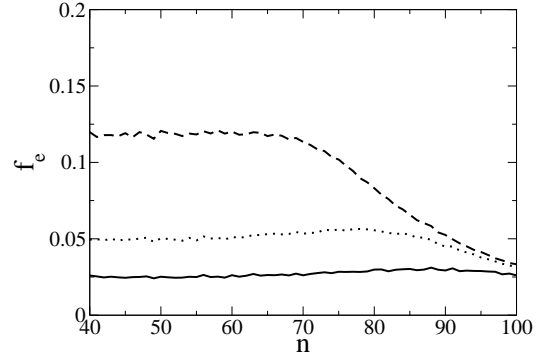}
       \caption{The fraction of excited Rydberg atoms $f_{e}$ as a function of 
       increasing excitation $n$ and for different laser pulse length 
       of 1$\mu$s (solid), 2$\mu$s (dotted), and 5$\mu$s (dashed). The density of the ultracold gas is 
       $\rho = 8\cdot 10^{9}$cm$^{-3}$, and  the parameters are close to those of the 
       experiment \cite{sire+04},
	$\Omega=2\pi\cdot 22.1\,$MHz, $\omega=2\pi\cdot 
	0.8\,$MHz, 
	$\Gamma = 2\pi\cdot 6\,$MHz.}
       \label{fig:blockade}
   \end{figure}

However, we are interested in the many-atom case of an ultracold gas, 
where the laser is tuned to the single atom resonance and the 
detuning $\Delta$ is given by an interaction induced shift $\delta$ 
of $|e\rangle$.
 Since the shift is additive it is easily generalized to this
 case where now  $\delta_{i}$ for  atom $i$ is given by the
 sum $\delta_{i}=\sum_{j}\delta(|\vec r_{i}-\vec r_{j}|)$, with the
 index $j$ running over all Rydberg atoms existing at this time and
 $\delta(R)$ given in \eq{shift}.  Number and location of Rydberg
 atoms at a time can be determined by solving the rate equation
 \eq{reexc} for each atom with a Monte Carlo technique.  The result is
 shown in \fig{fig:blockade}, namely a decreasing fraction of Rydberg
 atoms with increasing level shift $\delta$, here realized through
 increasing excitation $n$. 
 This is the typical blockade effect,
 observed in \cite{sire+04,tofa+04}.
 
 The parameters of \fig{fig:blockade} are close to those of the
 experiment \cite{sire+04}.  However, for those parameters, the single
 atom excitation probability $P_{e}(t,\Delta)$ differs qualitatively
 from the one shown on the left part of \fig{fig:noDP}.  It has a
 double peak structure due to an Autler-Townes splitting induced by
 the strong driving $\Omega$, as can be seen on the right part of
 \fig{fig:noDP}, with maxima at $\Delta = \pm \Delta_{m}$.
 Due to the wide distribution of mutual atomic distances $R$ in a gas
 the characteristic double peak excitation profile with a peak
 separation of $\Delta \approx \Omega$ does not have an effect on
 $f_{e}$ as shown in \fig{fig:blockade}.  
 
\begin{figure}[ht]
      \includegraphics[width=0.8\columnwidth]{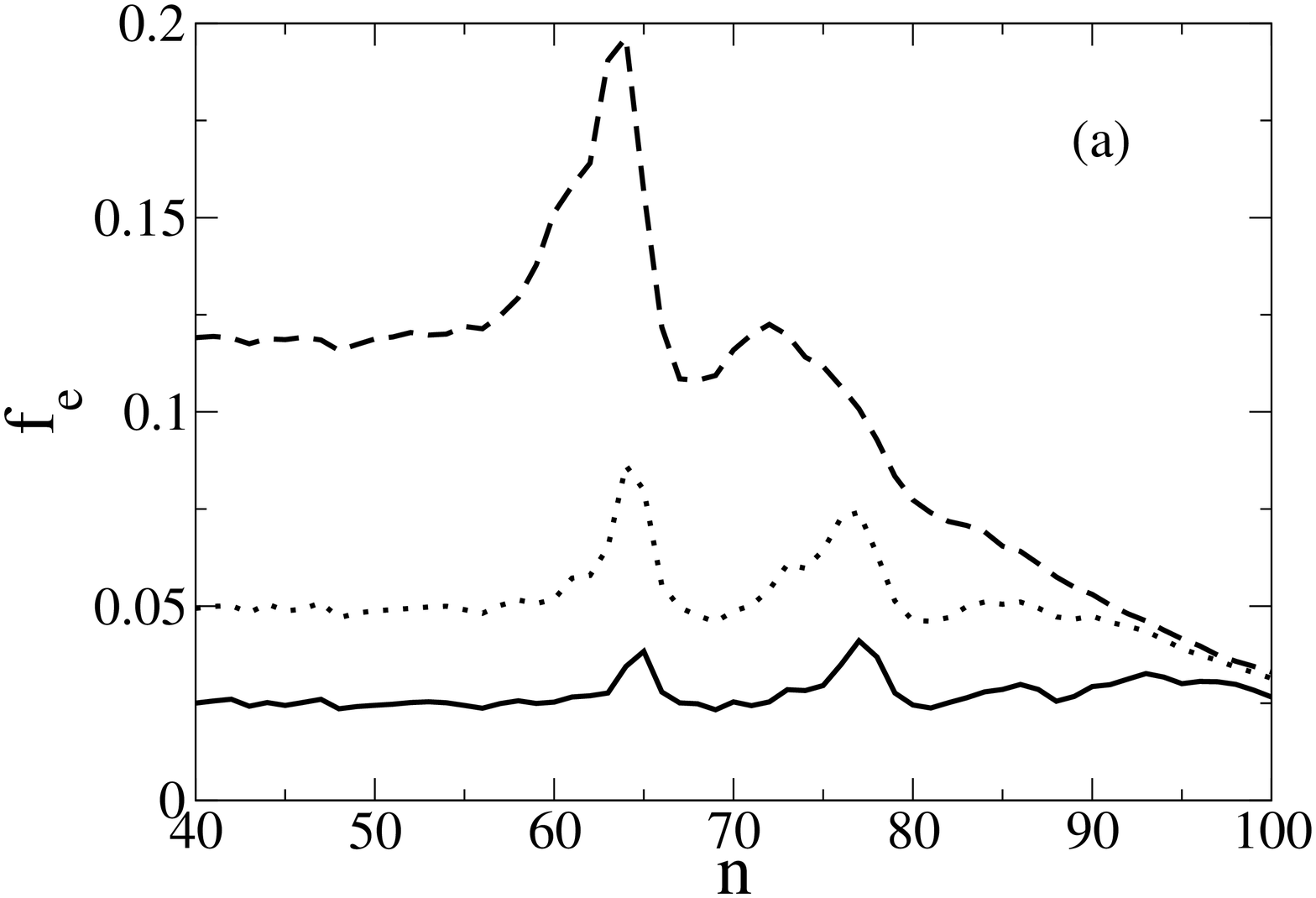}
      
      \includegraphics[width=0.8\columnwidth]{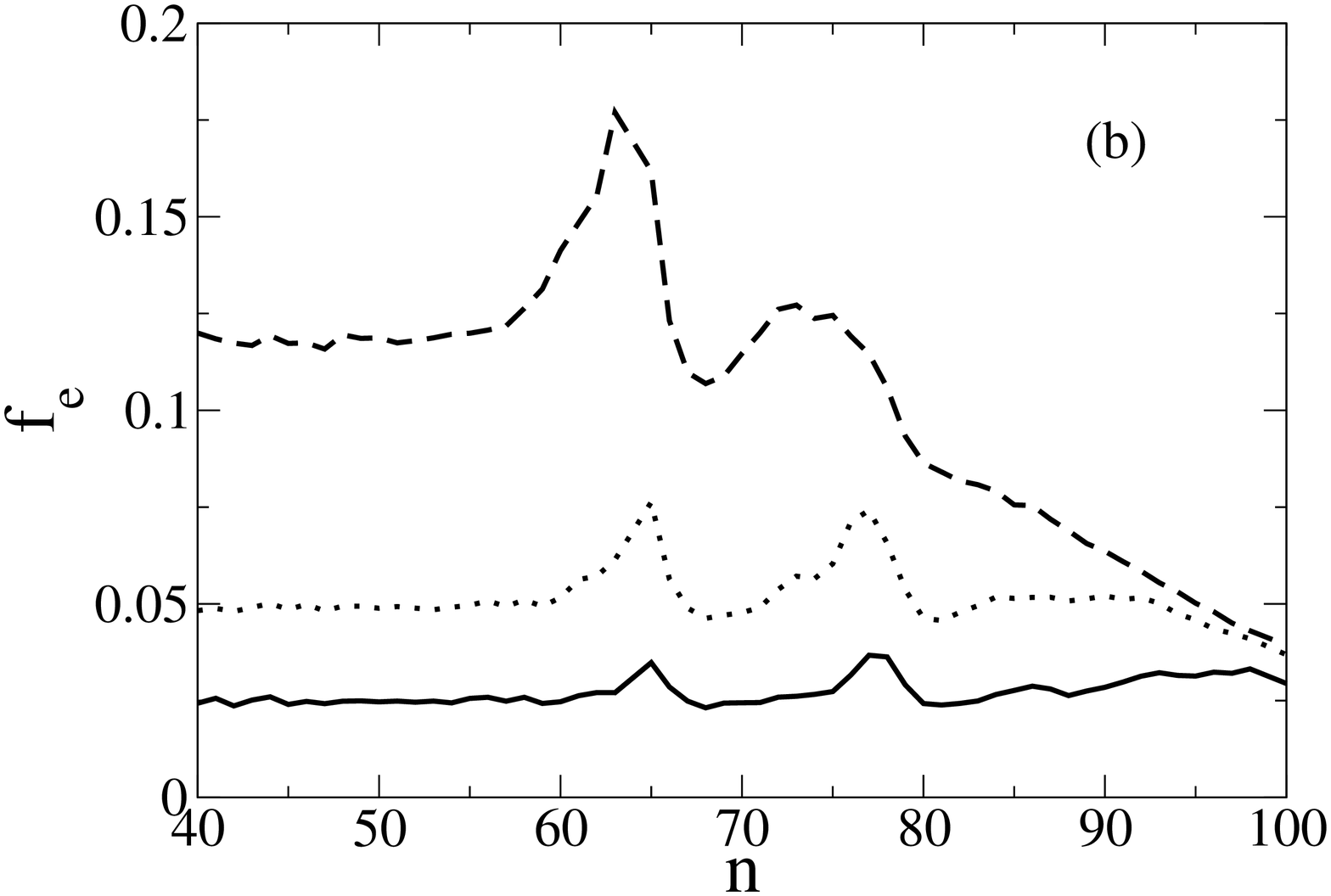}
      \caption{
       The fraction of excited Rydberg atoms $f_{e}$  as a function of 
      increasing excitation $n$  for atoms  regularly arranged in a
      3-dimensional simple cubic lattice with lattice constant $5\mu$m 
      ($\rho = 8\cdot 10^{9}$cm$^{-3}$) with parameters as in 
      \fig{fig:blockade};
      (a) perfect filling of lattice sites, (b)
      20\% lattice defects (i.e., empty lattice sites).}
      \label{fig:antiblockade}
  \end{figure}

\begin{figure}
      \includegraphics[width=0.5\columnwidth]{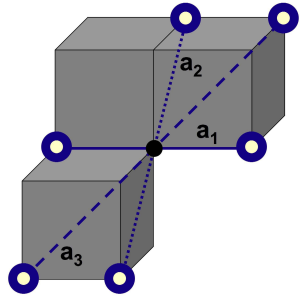}
      \caption{Arrangement of Rydberg and ground state atoms leading
      to antiblockade peaks and satellite peaks.}
      \label{fig:lattice}
  \end{figure}

  To make the effect of the double peak in the excitation profile
  visible, one needs a situation where the distribution of mutual
  atomic distances of the atom is restricted.  This can be achieved,
  e.g., with a lattice gas, i.e., with atoms regularly spaced on a
  three dimensional lattice.  The fraction of excited atoms on a 
  simple cubic
  lattice with a lattice constant $a=5\mu m$ is shown in
  \fig{fig:antiblockade}.
 One can see the known Rydberg blockade but on top of it an
 antiblockade effect, i.e., an enhanced fraction of excited atoms for
 certain values of $n$.  A closer look at \fig{fig:antiblockade}
 reveals that there are even satellite peaks next to the main peaks.
 This pattern of peaks is easily understood if one analyzes the
 geometry of the atoms placed on a cubic lattice (see
 \fig{fig:lattice}).  The shift  
 in the Rydberg level of a ground state atom
 is dominantly induced by a Rydberg atom  located
 on the next lattice site at a distance of the lattice constant  $a$.
 Hence we may write for the shift $\delta(a;n)$. For a certain $n$
 of the Rydberg atom
 this shift matches the  detuning $\Delta_{m}$ needed
 for maximum excitation rate (see right part of \fig{fig:noDP}),
 which leads to a peak in
 \fig{fig:antiblockade}. As experimentally also done, by changing the 
 laser intensity we can reach different $n$ keeping the Rabi 
 frequency $\omega$ constant.
  Clearly, with different $n_{i}$ the
 (optimum) shift $\Delta_{m}$ can be achieved by a Rydberg atom at
 a different distance $a_{i}$, so
 that in general 
 \be{condition} \delta(a_{i};n_{i})=\Delta_{m}\,.  
 \ee 
 The obvious
 candidates in a cubic lattice apart from the nearest neighbor at
 $a_{1}=a$ are the neighbors along the diagonal at $a_{2}=\sqrt 2a$
 and along the cubic diagonal at $a_{3}=\sqrt 3a$.  If one
 calculates the corresponding quantum numbers $n_{i}$ from \eq{condition},
 one would predict $\{n_{1},n_{2},n_{3}\}=\{65,78,87\}$.
 This differs at most by one quantum of excitation from 
 the  numerical values of 
 \fig{fig:blockade} which are $\{65,77,86\}$ for the shortest pulse 
 length ($1\mu$s).
 Of course, for the longer pulses the interaction is stronger with 
 a considerable amount of excited atoms. This  background of excited 
 atoms leads  to a general shift of the peaks towards lower $n$.

 In addition to the main peaks described, satellite peaks
 can be seen to the left of the main peaks.  
 They
 correspond to a situation where two Rydberg atoms on each side of a
 central atom contribute half  to the shift of the excited level of
 the central atom, see \fig{fig:lattice}. 
 Of course, for this enhancement less interaction
 is necessary than in the nearest neighbor enhancement.  Therefore these
 satellite peaks appear at smaller $n$, to the left of the main peaks. 
 Since double the
 number of Rydberg atoms is necessary,  the satellite peaks are 
 roughly
  only half the height of the main peaks in the linear regime of low concentration of 
  Rydberg atoms (short laser pulse, solid curve in \fig{fig:blockade}). 
 
\begin{figure}
    \includegraphics[width=0.8\columnwidth]{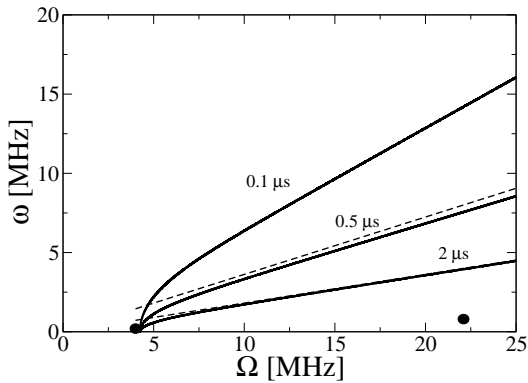}
    \caption{Phase diagram for $P_{e}(t,\Delta)$ from \eq{reexc}.
    Solid lines represent the phase boundaries between the blockade
    (upper area) and the antiblockade (lower area) regime for different
    pulse lengths, while $\Gamma=2\pi\cdot 6$Mhz. The dashed lines are the linear approximations for
    $\Omega\gg\omega$, see text.  The parameter sets used in
    \fig{fig:noDP} are marked with black dots.}
    \label{fig:phasediag}
\end{figure}

 A lattice with characteristic spacing of $a\sim 5\mu$m can be 
 produced by CO$_{2}$ lasers \cite{frwa+98}, but probably can only be filled leaving 
 a significant amount of lattice defects. Therefore, we have 
 simulated a situation with 20\% random lattice defects. 
 The defects hardly diminish the contrast of the 
 antiblockade effect in the fraction of excited atoms 
 (\fig{fig:antiblockade}b),  since the antiblockade relies dominantly 
 on contributions 
 from pairs of close neighbors: If a neighbor is missing due to a 
 lattice defect this does 
 neither lead to antiblockade nor to blockade and therefore does not 
 spoil $f_{e}$.
 The large tolerance with respect to lattice 
 defects makes an experimental verification of  the antiblockade 
 feasible, depending on the appropriate combination of 
 Rabi frequencies, pulse length and decay rate $\Gamma$.
 The corresponding ``phase diagram'' with the boundary between the 
 single peak (blockade) and double peak (antiblockade) domain is 
 shown in \fig{fig:phasediag}. The rate equation \eq{reexc} reveals 
 that for large $\gamma t/P_{\infty}$, the exponential is suppressed and the 
 steady state probability $P_{\infty}$ dominates, giving rise to a 
 single peak structure. On the other hand, for small $\gamma 
 t/P_{\infty}$ the Rydberg population is governed by $\gamma$ which 
 gives rise to a double peak structure. 
 The rate $\gamma$ as well as $P_{\infty}$ take a 
 relatively simple form in the limit $\omega\ll\Gamma\ll\Omega$,
 namely
 \be{coeffs}
  \gamma=\frac{\Gamma\omega^{2}/\Omega^{2}}
 {2(1-4\Delta^{2}/\Omega^{2})^{2}}\,\,\,\,\,\,\,\,\,
 P_{\infty}=\frac{1}{1+8\Delta^{2}/\Omega^{2}}\,.
 \ee
 The coefficients \eq{coeffs} reveal a universal condition for the
 transition from the double to the single peak structure of
 $P_{e}(t,\Delta)$, defined by
 $\partial^{2}P_{e}(t,\Delta)/\partial\Delta^{2}|_{\Delta=0}=0$ which
 can be written as
 \be{g0}
 g_{0} = 2\ln(1+g_{0})
 \ee
 with  $g_{0}=\Gamma
 t\omega^{2}/\Omega^{2}$. Equation (\ref{g0}) is easily solved by 
 iteration to give $g_{0}=2.513$. Hence, for $\Omega\gg\omega$ we expect a
 linear line $\omega = \alpha \Omega$ separating the two regimes,
 where $\alpha^{2}=g_{0}/(t\Gamma)$ which is indeed the case 
 (dashed lines in
 \fig{fig:phasediag}).
 Note also, that \eq{coeffs} clearly demonstrates the transient 
 character of the double peak structure which vanishes for long laser 
 pulses. Yet,  experimentally accessible  parameters, e.g.,  in 
 \cite{sire+04},
 realize exactly the transient regime and therefore provide the conditions
 to see the antiblockade.

To summarize, we have derived a rate equation for the
population of Rydberg states in ultracold gases by a two-step excitation
process.  The rate describes very well the structure of the Rydberg
excitation in a single atom when compared to the exact solution of the
Bloch equations including a non-trivial transient Autler-Townes splitting in the
Rydberg population for certain parameters.

The validity of the rate equation has allowed us to formulate the
many-body excitation dynamics of Rydberg states in an ultracold gas
without a mean-field approximation \cite{tofa+04} and for a realistic
number of atoms \footnote{Even for a small number of atoms one can
solve the problem hardly quantum mechanically, see \cite{rohe05}.} as a
random Monte Carlo process. We can reproduce the observed Rydberg
blockade effect observed previously and also its effect on the atom 
counting statistics \cite{atpo+06} as measured in \cite{lire+05},
but in addition we have identified an antiblockade effect due to the 
Autler-Townes splitted Rydberg population.  
We predict that this antiblockade effect can be seen in an experiment 
with a gas trapped in an optical lattice created by a CO$_{2}$ laser 
since the antiblockade effect is robust even against a large number 
of lattice defects. In the (realistic) limit of a very small upper Rabi 
frequency $\omega$ we could show that the formation of the double or 
single peak structure in the Rydberg population is determined by a 
universal parameter which allows a simple navigation in parameter 
space consisting of the two Rabi frequencies, the decay rate of the 
intermediate level and the pulse length, to achieve the desired peak 
structure in the single-atom Rydberg excitation probability. 


\end{document}